\documentclass[12pt, a4paper]{amsart}

\usepackage[utf8]{inputenc}
\usepackage[T1]{fontenc}
\usepackage{lmodern}
\usepackage{amsmath,amssymb} 
\usepackage{graphicx, xcolor}
\usepackage{booktabs}
\usepackage{xurl}            
\usepackage[margin=1in]{geometry} 
\usepackage{amsthm}
\usepackage{physics}
\usepackage{comment}
\usepackage{microtype} 

\usepackage{tcolorbox}
\usepackage{tikz}

\usepackage{hyperref}       

\usepackage[
  backend=biber,             
  citestyle=authoryear,      
  bibstyle=numeric,          
  maxcitenames=2,            
  maxbibnames=99             
]{biblatex}

\usepackage[dvipsnames]{xcolor}
\usepackage{enumitem}

\setlist[enumerate]{
  before=\vspace{0.5ex},
  after=\vspace{0.5ex},
  topsep=0.1ex,
  itemsep=0.5ex,
  parsep=0ex,
}

\setlist[itemize]{
  before=\vspace{0.5ex},
  after=\vspace{0.5ex},
  topsep=0.1ex,
  itemsep=1ex,
  parsep=0ex,
}

\makeatletter
\renewcommand{\subsection}{\@startsection{subsection}{2}{\z@}%
  {-3.25ex\@plus -1ex \@minus -.2ex}%
  {1.5ex \@plus .2ex}%
  {\normalfont\normalsize\bfseries}} 
\makeatother

\makeatletter
\renewcommand{\subsubsection}{\@startsection{subsubsection}{3}{\z@}%
  {-3.25ex\@plus -1ex \@minus -.2ex}%
  {1.5ex \@plus .2ex}%
  {\normalfont\normalsize\bfseries}} 
\makeatother

\makeatletter
\renewcommand{\paragraph}{\@startsection{paragraph}{4}{\z@}%
  {3.25ex \@plus 1ex \@minus .2ex}%
  {0.5ex}
  {\normalfont\normalsize\bfseries}}
\makeatother

\definecolor{light-gray}{gray}{0.95}

\selectfont

\title[Managing Ambiguity: A Proof of Concept of Human-AI Symbiotic Sensemaking]{Managing Ambiguity: A Proof of Concept of Human-AI Symbiotic Sensemaking based on Quantum-Inspired Cognitive Mechanism of Rogue Variable Detection$^{1}$}

\author[A. Bieńkowska]{Agnieszka Bieńkowska}
\address{Department of Management Systems and Organizational Development, Faculty of Management, Wrocław University of Science and Technology}
\email{agnieszka.bienkowska@pwr.edu.pl}

\author[J. Małecki]{Jacek Małecki}
\address{Department of Mathematics, Faculty of Mathematics, Wrocław University of Science and Technology}
\email{jacek.malecki@pwr.edu.pl}

\author[A. Mathiesen-Ohman]{Alexander Mathiesen-Ohman}
\address{AMOTHO Research Institute, Vallsjön 20, 780 00 Rörbäcksnäs, Sweden}
\email{amo@amotho.com}

\author[K. Tworek]{Katarzyna Tworek}
\address{Department of Management Systems and Organizational Development, Faculty of Management, Wrocław University of Science and Technology}
\email{katarzyna.tworek@pwr.edu.pl}

\dedicatory{\today}


\begin{document}

\begin{abstract}
    Organizations increasingly operate in environments characterized by volatility, uncertainty, complexity, and ambiguity (VUCA), where early indicators of change often emerge as weak, fragmented signals. Although artificial intelligence (AI) is widely used to support managerial decision-making, most AI-based systems remain optimized for prediction and resolution, leading to premature interpretive closure under conditions of high ambiguity. This creates a gap in management science regarding how human-AI systems can responsibly manage ambiguity before it crystallizes into error or crisis.

    This study addresses this gap by presenting a proof of concept (PoC) of the LAIZA human-AI augmented symbiotic intelligence system and its patented process: Systems and Methods for Quantum-Inspired Rogue Variable Modeling (QRVM), Human-in-the-Loop Decoherence, and Collective Cognitive Inference. The mechanism operationalizes ambiguity as a non-collapsed cognitive state, detects persistent interpretive breakdowns (rogue variables), and activates structured human-in-the-loop clarification when autonomous inference becomes unreliable.

    Empirically, the article draws on a three-month case study conducted in 2025 within the AI development, involving prolonged ambiguity surrounding employee intentions and intellectual property boundaries. The findings show that preserving interpretive plurality enabled early scenario-based preparation, including proactive patent protection, allowing decisive and disruption-free action once ambiguity collapsed. The study contributes to management theory by reframing ambiguity as a first-class construct and demonstrates the practical value of human-AI symbiosis for organizational resilience in VUCA environments.

\end{abstract}

\footnotetext[1]{\textbf{Alexander Mathiesen-Ohman, \textit{Systems and Methods for Quantum-Inspired Rogue Variable Modelling (QRVM), Human-in-the-Loop Decoherence, and Collective Cognitive Inference in Human-AI Symbiotic Architectures}, Provisional Utility Patent No 63/941,301}}

\maketitle

\section{Introduction}

Modern organizations increasingly operate in environments (both internal and external) characterized by volatility, uncertainty, complexity, and ambiguity (VUCA) (\cite{JohansenEuchner2013}). Originally developed in military and strategic research, and subsequently adopted in management studies, the VUCA framework reflects conditions in which traditional mechanisms of planning, forecasting, and control within an organization lose their effectiveness. In such environments, decision-makers are faced not only with a lack of information relevant to the decisions they make (despite its abundance in general terms, which may be a consequence of, for example, an inability to detect weak signals), but rather with its high volatility and, at times, its ambiguity, including rapidly shifting, conflicting, and weakly structured signals that resist straightforward interpretation. The term "signals" should be understood as time-varying events, words, messages, stimuli, emotions, behaviors, as well as preferences or intentions that carry information, enabling communication and understanding of the state of reality, and thus influencing decisions made in the organization, not only of a strategic nature, but also tactical or operational. In the decision-making process, those signals need to be (1) identified, to be (2) interpreted, and based on it (3) used for the final decision making. One of the specific categories of signals important for the functioning of an organization are weak signals, which are evidence-based, early, imprecise, and therefore difficult to perceive, as well as surprising, uncertain, irrational, or even unreliable signs of inevitably approaching significant events and their consequences (very strong phenomena) (\cite{Ansoff1975}). Weak signals are not only difficult to detect and pick out from the "information noise," but also cause interpretation problems due to, among other things, their imprecision and uncertainty. It should be emphasized, however, that it is not only weak signals that can be characterized by uncertainty or ambiguity.

\medskip

At the same time, organizations increasingly rely on hybrid decision systems in which humans and artificial intelligence (AI) jointly shape actions, forecasts, and strategic choices. These systems are currently present in strategic management, leadership support, human resource decisions, risk assessment, and organizational analysis. Although AI-assisted systems are excellent at identifying factors that influence the functioning of an organization and processing large amounts of data (including even those that may be weak signals) and identifying stable and universal patterns, even assuming dynamic variability within them, this is not sufficient for modern organizations. AI remains optimized for broad and rapid access to knowledge bases, which is indeed the basis for forecasting and problem solving, as it enables rapid detection of even weak signals, especially those coming from the organization's environment, but it does not fully secure the decision-making process. Admittedly, progress in this area is dynamic, access to knowledge is increasingly broad and error-free, but unfortunately even this is not sufficient for decision-making (especially in managing the external and internal VUCA environment), because, for example, not all AI systems are equipped with the ability to match the information provided by AI to the changing needs and context relevant to the user (P-AI fit, dynamic P-AI fit (\cite{BienkowskaMMT2025}). So, while signals from the environment are within the reach of AI systems, data from within the organization are not. What's more, unfortunately, AI systems remain vulnerable to a fundamental limitation: ambiguity in VUCA conditions.

\medskip

Situation, in which ambiguity of signal arises occurs when signals relevant to decision-making are incomplete, contradictory, unstable, or transitional, such that no single interpretation can be confidently selected (\cite{March1978}; \cite{Weick1995}; \cite{Kail2011}). The ambiguity of signals makes interpretation difficult (or even impossible) in this context, as it is directly related to assigning meaning to these signals in a specific context. Contextual meaning-making should be understood as a dynamic process of interpretation, in particular psychological and social process of attributing meaning to the aforementioned signals, which shapes the perception of reality and influences decisions, often through the prism of personal experiences, values, and context. In this context, it should be emphasized that assigning meaning to signals depends on the person who assigns that meaning (the observer) and the situation in which they find themselves.

\medskip

In VUCA environments (both internal and external), ambiguity is not an exception but a persistent state, frequently preceding critical organizational events such as strategic inflection points, leadership stress and cognitive overload, ethical dilemmas, employee disengagement or burnout, and cascading coordination failures. They therefore require effective detection and correct interpretation, taking into account the dynamics of phenomena, preceding the decision-making process and significantly influencing its shape. Importantly, these situations often emerge before any observable behavioral collapse or measurable performance decline, making them particularly difficult to detect (due to the lack of noticeable results from these situations) and manage using conventional analytical approaches.

\medskip

One approach that is particularly important for detecting and identifying situations of ambiguity is the theory of weak signals, which refers to the weak signals described above, and also allows to consider subtle, early, and often ambiguous indicators of emerging change (\cite{Ansoff1975}). Weak signals are typically fragmented, low-intensity, and difficult to distinguish from information noise. Yet they often precede major disruptions, crises, or strategic shifts. Effective managerial sensemaking in VUCA environments depends not on eliminating ambiguity, but on recognizing and engaging with occurring signals (including weak signals) before they consolidate into strong, irreversible outcomes (\cite{WeickSutcliffeObstfeld2005}). 

\medskip

Most modern artificial intelligence (AI) systems are capable of detecting signals that are relevant to organizational management. However, they are optimized for interpretive resolution, which means quickly assigning meaning (sense) to a given signal rather than suspending interpretation. Faced with ambiguous or weak signals, such systems tend to prematurely converge on a single inferred state, recommendation, or forecast. While computationally efficient, this tendency can suppress early warning signs, amplify model-driven bias, and reduce organizational sensitivity to emerging threats and opportunities. In internal and external VUCA conditions, such premature closure may increase risk of making wrong decision rather than mitigate it.

\medskip

This tension exposes a persistent theoretical and practical gap in management science: how AI-based systems can detect, represent, and responsibly support organizational management in conditions of ambiguity before it turns into errors, discrepancies, or harmful actions that negatively impact the reliability of the organization as a whole. While prior research has addressed decision quality, explainability, algorithmic bias, and human–AI collaboration, comparatively little attention has been paid to pre-decisional cognitive states—states in which meaning remains unstable, interpretations coexist, and weak signals have not yet been resolved.

\medskip

This article addresses this gap by introducing and empirically examining an innovative solution developed within the H3LIX Project: a quantum-state–based cognitive mechanism for rogue variable detection, protected under patent number $63/941,301$ (\cite{Mathiesen2025b}). The mechanism is implemented in LAIZA, a human–AI augmented symbiotic intelligence system (\cite{Mathiesen2025a}) designed to support managerial cognition, decision-making and creating organizational meaning in conditions of internal and external VUCA, and ambiguity of signals that are important from the point of view of the organization's functioning.

\medskip

Rather than forcing early classification or prediction, the H3LIX Project mechanism operationalizes ambiguity as a detectable and manageable cognitive condition. It introduces the concept of rogue variables—configurations of cognitive, behavioral, and contextual signals that indicate a breakdown in interpretive coherence and signal that existing cognitive or organizational models are insufficient. Rogue variables often correspond to weak signals: early manifestations of stress, misalignment, or emerging change that are not yet actionable in conventional analytical terms.

\medskip

The H3LIX/LAIZA Human-AI augmented symbiosis system, presented in this paper, employs a quantum-inspired state representation that allows multiple competing interpretations to coexist without premature resolution. When internal inference reaches its limits, the mechanism activates a structured human-in-the-loop clarification process, ensuring that ambiguity is resolved through human sensemaking rather than automated extrapolation. In this way, H3LIX/LAIZA functions not as a decision replacement system, but as a sensemaking amplifier, one explicitly designed for VUCA environments, ethical governance, and responsible human–AI interaction. It should be emphasized that the adopted solution, and thus the essential thought process for decision-making, is carried out by humans, which, apart from giving them a fairly obvious freedom of interpretation and responsibility for the outcome, also forces them to make a mental effort, which seems to be the essence of humanity according to the principle of "cogito ergo sum."

\medskip

This article makes two primary contributions. First, it advances management theory by integrating VUCA theory, weak signals theory, and human–AI symbiosis, conceptualizing ambiguity as a first-class construct rather than a residual problem to be eliminated. Second, it presents a case study demonstrating the efficiency and practical value of the H3LIX project rogue variable detection mechanism as implemented in H3LIX/LAIZA (\cite{Mathiesen2025b}). The case study illustrates how the system enhances early detection of weak signals, reduces premature decision closure, and supports more resilient human–AI collaboration in complex organizational settings.

\medskip

By bridging established theories of VUCA and weak signals with a concrete, patent-backed cognitive mechanism, this study contributes to ongoing debates on responsible AI, organizational resilience, and the future of management systems operating under persistent uncertainty.

\section{Ambiguity in VUCA environment and weak-signals theory}

Contemporary organizations increasingly operate within internal and external environments characterized by volatility, uncertainty, complexity, and ambiguity (VUCA) - a condition that reflects the structural transformation of economic, technological, and social systems (\cite{JohansenEuchner2013}). Volatility refers to the speed and magnitude of change, uncertainty to the limited predictability of future events, complexity to the interdependence of multiple interacting factors, and ambiguity to the absence of clear cause–effect relationships. Unlike earlier periods dominated by relatively stable competitive dynamics and linear planning assumptions, today’s organizational environments are shaped by rapid technological innovation, digitalization, globalization, geopolitical instability, regulatory fragmentation, and accelerating societal change. These forces interact nonlinearly, producing outcomes that are difficult to forecast even with extensive data and analytical capabilities. As a result, organizations are frequently required to make strategic and operational decisions in situations where information is incomplete, signals are contradictory, and historical experience offers limited guidance. In such contexts, managerial effectiveness depends less on optimization and prediction and more on adaptive sensemaking, early signal recognition, and the ability to navigate ambiguity, making VUCA not a temporary disruption but a persistent condition of contemporary organizational life (\cite{Kail2011}).

\medskip

Organizations operations within VUCA environment, especially decision-making processes, require specific approach, which allows decision-makers to account for high ambiguity (Lawrence, 2013). One of those approaches is based on weak signal theory. Weak signal theory is a conceptual framework that helps organizations identify and act upon early indicators of potential crises or significant changes in their environment, managing „startegic surprise” (Ansoff, 1975). In the context of management science, it is particularly valuable for enhancing resilience and adapting strategies in volatile, uncertain, complex, and ambiguous (VUCA) environments. VUCA conditions present extraordinary challenges, making proactive management essential for survival and success.

\medskip

The framework of weak signals assists managers in recognizing subtle, often overlooked signs that may precede disruptive events (\cite{HolopainenToivonen2012}). Identifying these precursors allows organizations to intervene early, potentially preventing crises before they materialize. This approach is supported by literature arguing that effective crisis management can evolve from a predominantly reactive stance to a more proactive paradigm through the incorporation of weak signal identification (\cite{Wang2022}). By leveraging weak signals alongside other management strategies, organizations can create adaptive frameworks that enhance their ability to cope with unforeseen challenges (\cite{Chauhan2024}).

\medskip

Moreover, the theoretical underpinnings of weak signals interconnect with concepts such as risk management and antifragility, wherein organizations gain from disorder and complexity, when it is handled properly and before crisis occurs (\cite{Abernikhina2025}). This is critical for developing strategies that not only mitigate risks but also exploit potential disruptions for innovative growth (\cite{MinciuEtAl2025}). The focus on interdisciplinary solutions acknowledges that complexities inherent in crises, and before crisis, often span various sectors, necessitating a multifaceted approach (\cite{Achoki2023}).

\medskip

Building on this theoretical foundation, weak signal theory can be incorporated into contemporary managerial decision-making by reframing decisions as iterative sensemaking processes rather than discrete choice events (\cite{WeickSutcliffeObstfeld2005}). In VUCA environments, managers rarely face fully formed problems; instead, they encounter fragmented cues, contradictory feedback, and early disturbances that lack immediate interpretive clarity. Signals, including weak signals, by definition, do not justify decisive action on their own. Their value lies in shaping managerial attention, guiding inquiry, and informing the timing and scope of intervention (\cite{WeickSutcliffeObstfeld2005}; \cite{HolopainenToivonen2012}). As such, effective decision-making increasingly depends on a manager’s ability to temporarily suspend interpretation closure, tolerate ambiguity, and maintain multiple competing interpretations until sufficient coherence emerges.

\medskip

From a practical perspective, incorporating signals emerging into decision-making in order to manage the unexpected requires mechanisms that support early detection, interpretive plurality, and controlled escalation (\cite{WeickSutcliffeObstfeld2005}). Managers must be able to distinguish between information noise and meaningful, especially early, indicators without relying solely on hindsight or intuition. Prior research suggests that weak signals become actionable not through their immediate clarity, but through pattern accumulation, contextual embedding, and reflective interpretation over time (\cite{Ansoff1975}; \cite{Hiltunen2008}). This shifts managerial practice away from linear forecasting toward adaptive monitoring and reflective learning, where small deviations, emotional cues, and subtle behavioral changes are treated as legitimate inputs to strategic deliberation.

\medskip

Recent studies further emphasize that weak signal–informed decision-making benefits from structured support systems capable of integrating heterogeneous data sources while preserving interpretive openness (\cite{Venemyr2024}). Without such support, managers face cognitive overload, confirmation bias, and premature convergence—risks that are amplified under high pressure and time constraints. Consequently, weak signal theory increasingly intersects with research on organizational mindfulness, high-reliability organizing, and antifragility, which collectively argue that organizations can strengthen resilience by learning from early instability rather than suppressing it (\cite{WeickSutcliffeObstfeld2005}; \cite{Taleb2012}, \cite{Abernikhina2025}). In this view, weak signals are not merely warnings but opportunities for anticipatory adaptation and innovation, provided they are recognized and interpreted before crystallizing into crisis in situations with ambiguity.

\section{Quantum State Based Cognitive Mechanism of Rogue Variable Detection}

Most organizational decision models, both human and algorithmic, assume that cognitive states are stable enough to be classified (e.g., preference, intention, emotion, belief). Yet empirical research shows that many critical organizational moments occur before such stability emerges, where high level of ambiguity occurs. In the light of above mentioned weak signals theory, it seem to be important to indicate a mechanism which allows for the analysis of such ambiguities, which emerge in decision-making processes. 

\medskip

Such ambiguities are best understood as pre-determinate cognitive configurations, states in which multiple interpretations coexist and compete, without any one dominating.

\medskip

The proposed H3lix project framework, based on which H3LIX/LAIZA human-AI augmented symbiosis system (\cite{Mathiesen2025a}) operates, models ambiguity management as a five-stage socio-technical process embedded in a continuous human–AI symbiotic intelligence feedback loop, based on quantum state-based sognitive mechanism of Rogue Variable Detection (\cite{Mathiesen2025b}):

    \begin{enumerate}
        \item Cognitive State Representation
        \item Detection of Interpretive Breakdown (Rogue Variables)
        \item Human-in-the-Loop Clarification
        \item Organizational Memory of Ambiguity
        \item Cross-Individual Pattern Formation
    \end{enumerate}

Each stage is described below as a theoretical mechanism rather than a technical procedure.

\subsection{Representing Cognitive States Without Forcing Interpretation}

Cognitive states relevant to managerial decision-making are inherently latent and cannot be directly observed; they must instead be inferred from a constellation of heterogeneous signals. These signals typically span physiological cues (e.g., stress or arousal), behavioral patterns (such as interaction tempo or responsiveness), contextual conditions (including task demands and social environments), and internal system indicators reflecting uncertainty or instability in ongoing interpretations. Conventional decision-support systems tend to collapse such signals into singular inferred states, such as “high stress” or “low engagement”, thereby imposing premature coherence on situations that may remain fundamentally indeterminate. In contrast, the framework adopted in this study preserves these signals within a structured relational representation that explicitly maintains their tension and coexistence. By allowing multiple plausible interpretations to remain active simultaneously (\cite{Mathiesen2025b}), the H3lIX/LAIZA human-AI augmented symbiosis system aligns with core insights from management theory on sensemaking under equivocality (\cite{Weick1995}), bounded rationality, and paradox and ambidexterity research. Importantly, interpretation is deliberately deferred at this stage: ambiguity is treated not as a failure of analysis but as an analytically meaningful condition in its own right. This suspension of interpretive closure enables more reflective managerial engagement with emerging situations, particularly in environments characterized by volatility, uncertainty, complexity, and ambiguity.

\subsection{Detecting Rogue Variables as Interpretive Breakdown}

The second stage, H3LIX/LAIZA human-AI augmented symbiosis system focuses on the detection of rogue variables (\cite{Mathiesen2025b}), conceptualized as indicators of interpretive breakdown rather than data errors or missing information. A rogue variable emerges when the system’s internal expectations no longer align with observed cognitive, behavioral, or contextual signals, and when existing interpretive models prove insufficient to explain the evolving situation. Under such conditions, continued autonomous inference becomes unreliable or potentially unsafe. From an organizational perspective, rogue variables correspond to moments in which established mental models cease to function effectively, constituting situations in which decision-makers experience a sense that “something is off,” despite an inability to clearly articulate the source of concern. These moments often signal the early emergence of latent conflict, escalating risk, or impending disruption beneath the surface of routine organizational activity. Conceptually, rogue variables are identified through comparative pattern analysis that contrasts expected trajectories of cognitive and organizational states with their observed evolution over time. Rather than relying on isolated anomalies or fixed thresholds, the framework focuses on persistent mismatches and recurring configurations that generate interpretive instability. A condition is only recognized as a rogue variable when such instability endures across multiple contexts and temporal windows, thereby mirroring how experienced managers recognize deep ambiguity—not through singular events, but through sustained and unresolved divergence that calls for heightened attention and reflective sensemaking.

\subsection{Human-in-the-Loop Clarification as Ethical Control}

The third stage of the framework, based on which H3LIX/LAIZA human-AI augmented symbiosis system operates (\cite{Mathiesen2025b}), introduces human-in-the-loop clarification as a deliberate ethical control mechanism activated when a rogue variable is detected. At this point, the system intentionally suspends autonomous interpretation and decision-making, reflecting a central normative premise of the approach: ambiguity that cannot be resolved internally should be resolved by the human rather than the machine. This design choice aligns with principles of responsible AI, high-reliability organizing, and ethical decision-making in management, all of which emphasize restraint, accountability, and deference to human judgment under conditions of uncertainty. Rather than engaging in open-ended dialogue or attempting to infer intention, the system employs a tightly structured clarification process. It explicitly communicates the presence of ambiguity in neutral terms, poses a single minimal question aimed at resolving the specific interpretive gap, and avoids suggestion, persuasion, or prescriptive framing. For instance, instead of labeling physiological arousal as disengagement or stress, the system invites the human actor to distinguish between alternative interpretations, such as preparation for action or recovery from effort. By constraining interaction in this way, clarification becomes a focused sensemaking intervention rather than an intrusive interrogation, preserving human agency while enabling responsible resolution of ambiguity within the human–AI decision loop. Until clarification is received, all other predictions are suspended, recommendations are disabled, and autonomous actions are prohibited. From a management perspective, this models a form of deliberate organizational pause, analogous to escalation protocols in safety-critical industries.

\subsection{Organizational Memory of Ambiguity and Learning From Interpretive Experience}

The fourth step of H3LIX/LAIZA human-AI augmented symbiosis system in a given framework introduces an organizational memory mechanism designed to capture and retain experiences of ambiguity through a structured record of rogue variable episodes (\cite{Mathiesen2025b}). Each episode—whether ultimately resolved or left unresolved—preserves the configuration of ambiguity, the clarification process through which it was addressed, and the outcome of that intervention. Accumulated over time, these episodes form a longitudinal memory of ambiguous situations, enabling the system to recognize early warning signs, reduce unnecessary escalation of interpretive uncertainty, and progressively adapt expectations to individual and organizational baselines. This approach reframes organizational learning away from the optimization of discrete outcomes toward the accumulation of interpretive experience, particularly in relation to situations that were previously unclear, unstable, or weakly structured. In doing so, it operationalizes central insights from experiential learning theory, dynamic capabilities research, and organizational mindfulness, emphasizing learning from transitions, anomalies, and moments of uncertainty rather than solely from success or failure. By institutionalizing ambiguity as a legitimate source of learning, the framework supports more resilient and reflective forms of managerial sensemaking in VUCA environments.

\begin{figure}[htbp]
    \centering
    \includegraphics[width=\textwidth]{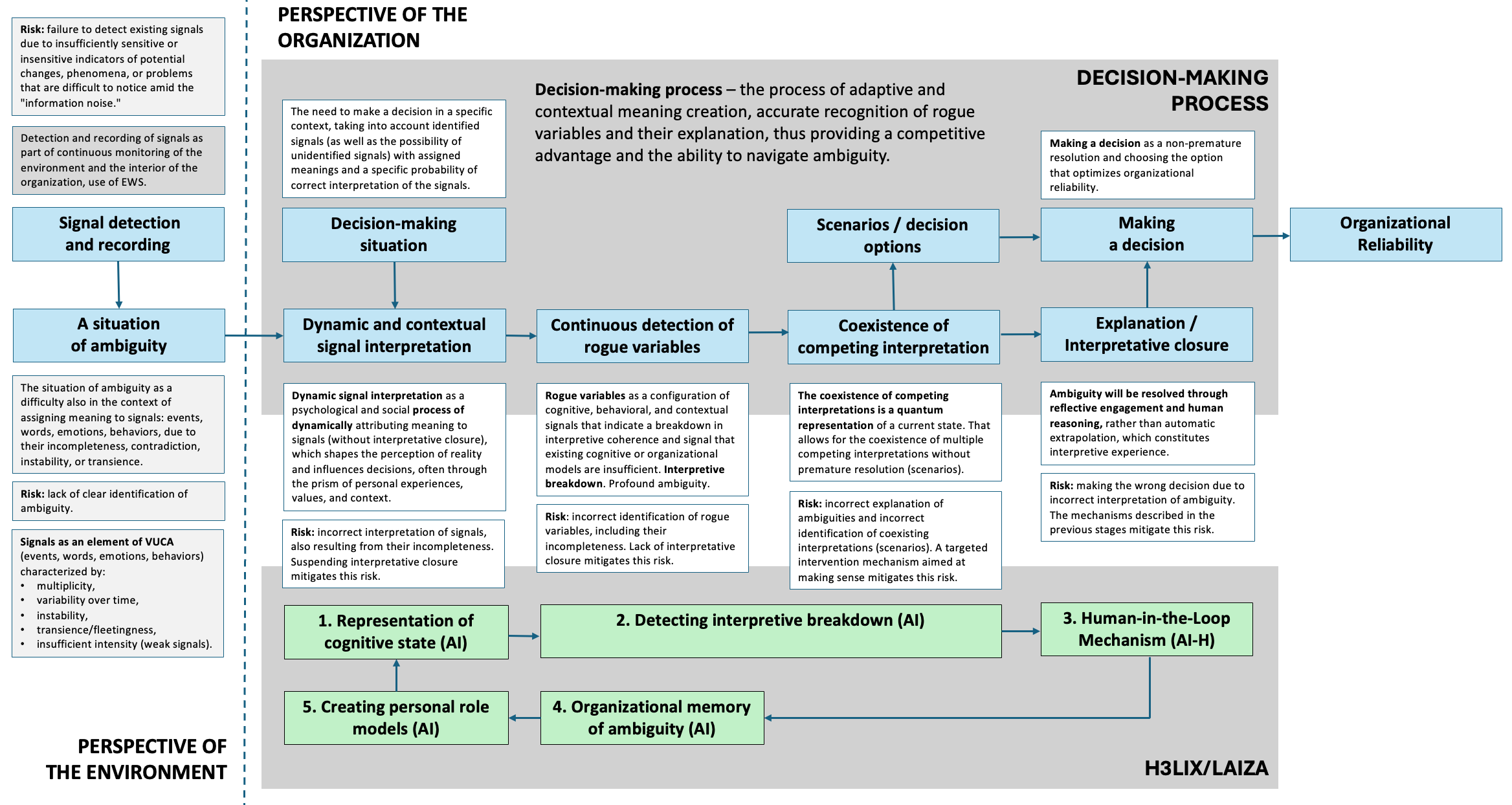}
    \caption{Decision-making process within the organization based on the Quantum State based cognitive mechanism of Rogue Variable Detection implemented in LAIZA human-AI augmented symbiosis system mechanism (own work)}
    \label{fig:decision_making_process}
\end{figure}

\subsection{Cross-Individual Pattern Formation and Collective Sensemaking}

The final stage of the framework for H3LIX/LAIZA human-AI augmented symbiosis system extends ambiguity management from the individual level to the collective domain by enabling the comparison of ambiguity patterns across multiple organizational actors while preserving individual privacy (\cite{Mathiesen2025b}). At this population level, recurring configurations of unresolved or emerging ambiguity can be identified, allowing the detection of shared stressors, collective cognitive overload, and early signs of organizational tension that may not be visible through individual analysis alone. By aggregating patterns of interpretive instability rather than final decisions or outcomes, the framework supports early identification of systemic risk and incipient crises before they manifest in observable breakdowns. Theoretically, this mechanism contributes to research on collective sensemaking, emotional contagion, and organizational climate by providing a processual account of how individual-level ambiguity accumulates and propagates across organizational systems. Importantly, it shifts analytical attention from post-hoc explanations of failure to the identification of pre-event patterns, offering a means to study how collective vulnerability and resilience emerge in real time within complex, VUCA environments.

\medskip

Based on the following description, Figure 1. Shows a decision-making process within the organization based on the Quantum State based cognitive mechanism of Rogue Variable Detection implemented in H3LIX/LAIZA human-AI augmented symbiosis system mechanism.

\section{Case Study: Managing Prolonged Ambiguity Through Scenario-Based Preparation in a Human–AI Development Organization}

\subsection{Case Context and Organizational Setting}

This case examines a three-month period in 2025 within the AI development Project, within organization operating at the intersection of advanced artificial intelligence development, research commercialization, and intellectual property creation. The focal situation concerned a senior employee holding a critical role as an AI technical lead. The employee was deeply involved in the development of core AI solutions that formed part of the organization’s strategic and technological foundation. Simultaneously, the employee was independently developing a side project situated within a closely related field of study and application.

\medskip

From the organization’s perspective, the situation was characterized not by overt misconduct or contractual breach, but signals were noticed that were impossible to interpret unambiguously, such as declining commitment (attribution of meaning: fatigue, too low pay, sabotage, desire to leave), failure to perform tasks according to the agreed schedule (attribution of meaning: desire to leave, lack of competence, external issues), discrepancies with external projects (attribution of meaning: fear of admitting mistakes, sabotage, desire to use the IP of the project). The likelihood of specific interpretations changed over time. There was also no certainty that all signals that occurred and were relevant in the context of the decision-making process were noticed. It caused a persistent ambiguity regarding future intentions. It remained unclear whether the employee planned to (a) continue contributing to the project while maintaining a separate but non-conflicting project, (b) seek eventual integration or acquisition of the side project, (c) appropriate intellectual property generated within the organization for personal use, or (d) disengage from the organization altogether. Complicating matters further, it was not evident whether potential IP-related risks stemmed from deliberate intent or from a lack of understanding of legal and organizational boundaries.

\subsection{Prolonged Ambiguity as a Non-Collapsed Cognitive State}

Throughout the three-month period, H3LIX/LAIZA represented the situation as a pre-determinate cognitive configuration, rather than forcing interpretive closure (\cite{Mathiesen2025b}). Behavioral indicators-such as continued technical delivery, participation in internal discussions, and ongoing system development-suggested engagement and professional commitment. At the same time, contextual and strategic signals pointed to increasing divergence: delayed resolution of IP-related discussions, guarded communication about long-term plans, and the independent maturation of the side project.

\medskip

Crucially, no single signal justified a definitive interpretation. The system therefore treated the situation as one of non-collapsed ambiguity, maintaining multiple plausible interpretations simultaneously. This approach aligned with weak signal theory and sensemaking under equivocality, allowing the organization to continue operations without escalation while remaining attentive to emerging risk.

\subsection{Rogue Variable Detection and Scenario Awareness}

As ambiguity persisted, H3LIX/LAIZA detected a rogue variable configuration reflecting sustained divergence between organizational expectations and observed patterns of behavior and communication. This detection did not frame the situation as an imminent crisis, but as an interpretive breakdown requiring heightened attention. Importantly, the system did not attempt to predict which scenario would materialize. Instead, it advised leadership to recognize that several mutually exclusive futures remained possible, none of which could yet be ruled out.

\medskip

This recognition marked a critical shift from decision-making toward scenario-based preparedness. Rather than choosing a single course of action prematurely, the organization was advised to remain operationally open while quietly preparing for multiple potential outcomes.

\subsection{Human-in-the-Loop Activation and Preparatory Strategy}

Human-in-the-loop interaction was activated once the accumulation of weak signals suggested a growing probability of employment termination, even though intent had not yet been explicitly stated. At this stage, executive leadership-supported by H3LIX/LAIZA’s interpretive guidance-confirmed that letting the employee go was a realistic possibility, though not yet a finalized decision.

\medskip

Because the state was still treated as ambiguous rather than resolved, H3LIX/LAIZA did not recommend confrontation or exclusion. Instead, it advised a parallel preparation strategy, including:

    \begin{itemize}
        \item continuing technical collaboration to avoid unnecessary disruption,
        \item initiating and advancing patent applications to protect intellectual property generated within the organization,
        \item seeking legal guidance regarding IP ownership, licensing, and potential acquisition of the side project,
        \item preparing exit scenarios that would allow for an orderly separation without reputational or operational damage.
    \end{itemize}

These actions were explicitly framed as preparatory rather than punitive. The system guide emphasized that ambiguity justified readiness, not accusation.

\subsection{Organizational Memory and the Value of Deferred Closure}

All preparatory steps were informed by the organizational memory of ambiguity maintained by the system. Because the situation had been recognized early as a rogue variable episode, actions such as patent drafting and legal consultation were initiated well before any definitive outcome was known. This allowed the organization to proceed methodically, without time pressure, emotional escalation, or reactive decision-making.

\medskip

In particular, the patent preparation process, which is typically time-consuming and resource-intensive-was able to progress under conditions of relative calm precisely because the system had resisted premature interpretive closure. Ambiguity was treated as a legitimate state requiring protection against downside risk, even in the absence of confirmed threat.

\subsection{Collapse of Ambiguity and Immediate Resolution}

The ambiguity collapsed when, during negotiations about the future of the employee, the individual explicitly stated his intention to leave the organization. At this moment, H3LIX/LAIZA was informed, via another iteration of Human-in-the-loop interaction, that the cognitive state had transitioned from ambiguity to resolution. Importantly, this collapse did not trigger the response to the crisis. Because preparatory actions had already been completed, no additional time was required to secure organizational interests.

\medskip

The patent protecting organizational IP was already in place, legal positions had been clarified, and exit procedures had been prepared. As a result, the employee was allowed to leave without repercussions, and the organization was able to proceed with strategic continuity and legal certainty.

\subsection{Case Implications}

This case demonstrates how treating ambiguity as a non-collapsed state enables organizations to prepare effectively for multiple futures without resorting to premature or adversarial action. By detecting weak signals, maintaining interpretive plurality, and activating human-in-the-loop clarification only when appropriate, the LAIZA system supported ethical, resilient, and strategically sound decision-making. The case illustrates that the primary value of the system lies not in prediction, but in temporal advantage—allowing organizations to act early under uncertainty so that, when clarity finally emerges, decisive action can be taken without delay or disruption.

\medskip

In this sense, the case provides empirical support for the argument that managing ambiguity proactively—rather than eliminating it prematurely—constitutes a critical capability for contemporary organizations operating in VUCA environments.

\begin{figure}[htbp]
    \centering
    s\includegraphics[width=\textwidth]{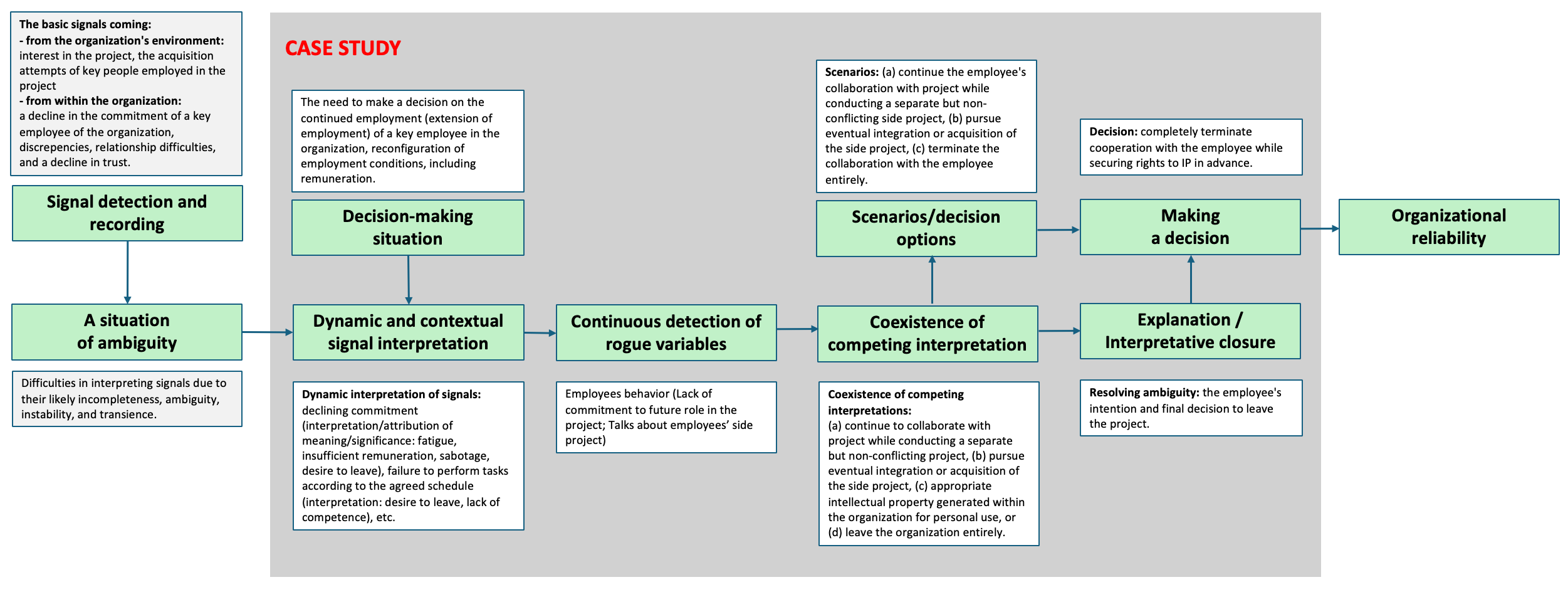}
    \caption{Case study decision-making process (own work)}
    \label{fig:case_study}
\end{figure}

\subsection{Formal Realization of the H3LIX/LAIZA Quantum Cognitive Mechanism}
\label{subsec:formal-realization-laiza}

This subsection documents the implementation of H3LIX/LAIZA in the proof-of-concept (PoC) case study and provides a concise account of how the underlying mathematical mechanism was operationalized during the three-month ambiguity episode.

\paragraph{Situation encoding as a time-indexed Mirrored Personal Graph (MPG).}
Throughout the period, H3LIX/LAIZA (\cite{Mathiesen2025a}; \cite{Mathiesen2025b}) maintained a time-indexed directed graph $G_t=(V_t,E_t)$ capturing the evolving configuration of signals relevant to the employee-intentions and IP-boundary ambiguity. Nodes $V_t$ represented managerially meaningful entities (e.g., delivery commitments and task progress, interaction patterns, openness about future plans, IP-boundary constructs, and organizational context such as milestones and legal/patent status). Directed edges $E_t$ represented relations among these entities (e.g., ``supports,'' ``contradicts,'' ``depends on,'' ``increases risk of,'' or ``temporally precedes''). Each node and edge carried feature vectors $m_t(v)$ and $m_t(e)$ aggregating heterogeneous inputs (behavioral/workflow cues, contextual cues, and internal model signals reflecting uncertainty and instability).

\paragraph{Quantum MPG State (QMS) as a representation of non-collapsed ambiguity.}
At each observation step, H3LIX/LAIZA computed a normalized state vector (\cite{Mathiesen2025b})
\begin{equation*}
\ket{\Psi_t}=\sum_{v_i\in V_t}\psi_t(v_i)\ket{v_i},
\qquad
\langle \Psi_t \mid \Psi_t \rangle = 1,
\label{eq:qms-case}
\end{equation*}
where amplitudes $\psi_t(v_i)$ were derived from node metrics and the current context. The activation weights $|\psi_t(v_i)|^2$ indicated which nodes were currently most relevant for explaining the situation. Crucially, this representation did not require assigning a single label such as ``disengaged'' or ``malicious intent.'' Instead, it allowed multiple plausible interpretations to remain simultaneously active. 


\begin{figure}[h!]
    \centering
    
    \begin{minipage}{0.49\textwidth}
        \centering
        \includegraphics[width=\linewidth]{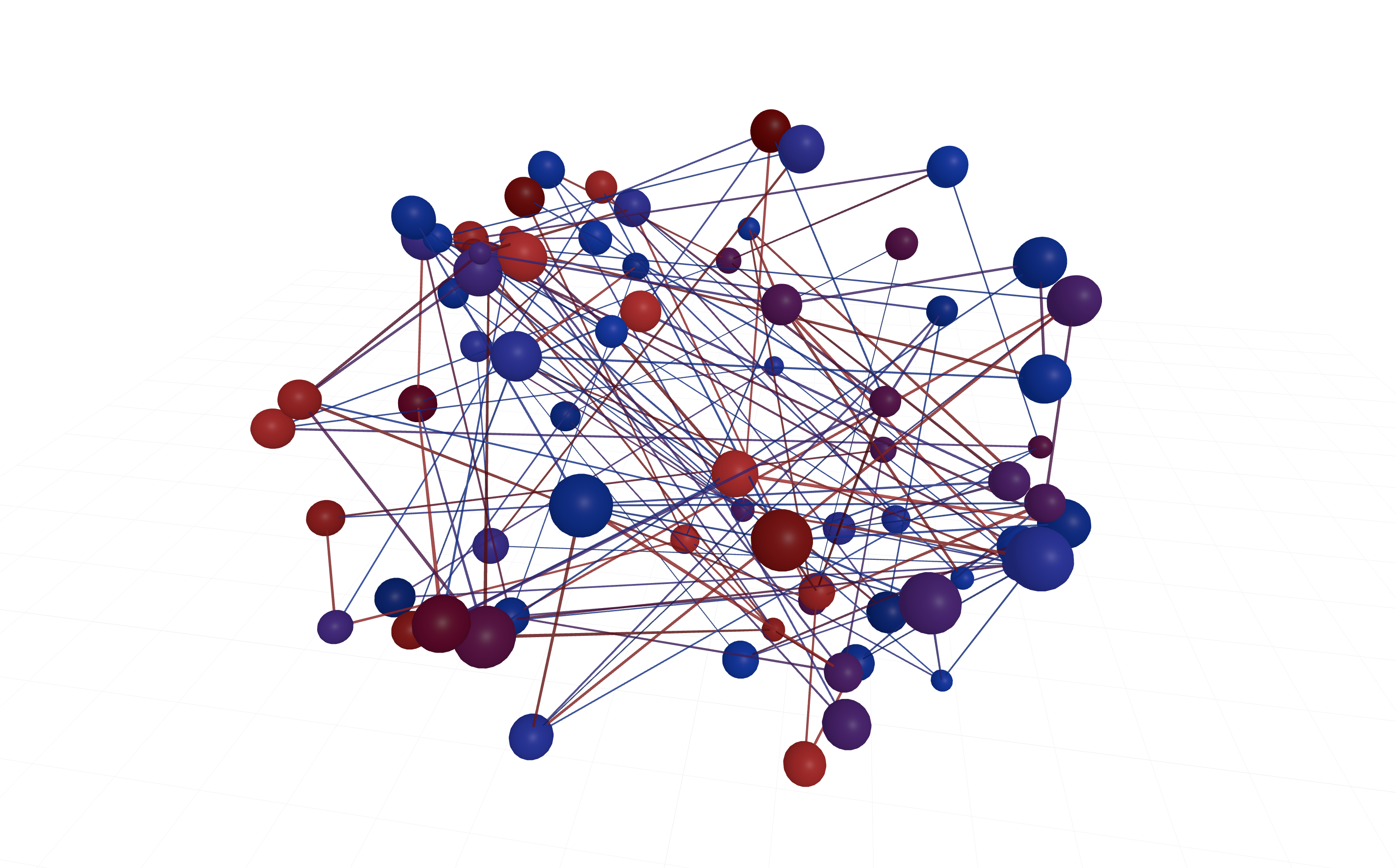}
    \end{minipage}%
    \hfill
    \begin{minipage}{0.49\textwidth}
        \centering
        \includegraphics[width=\linewidth]{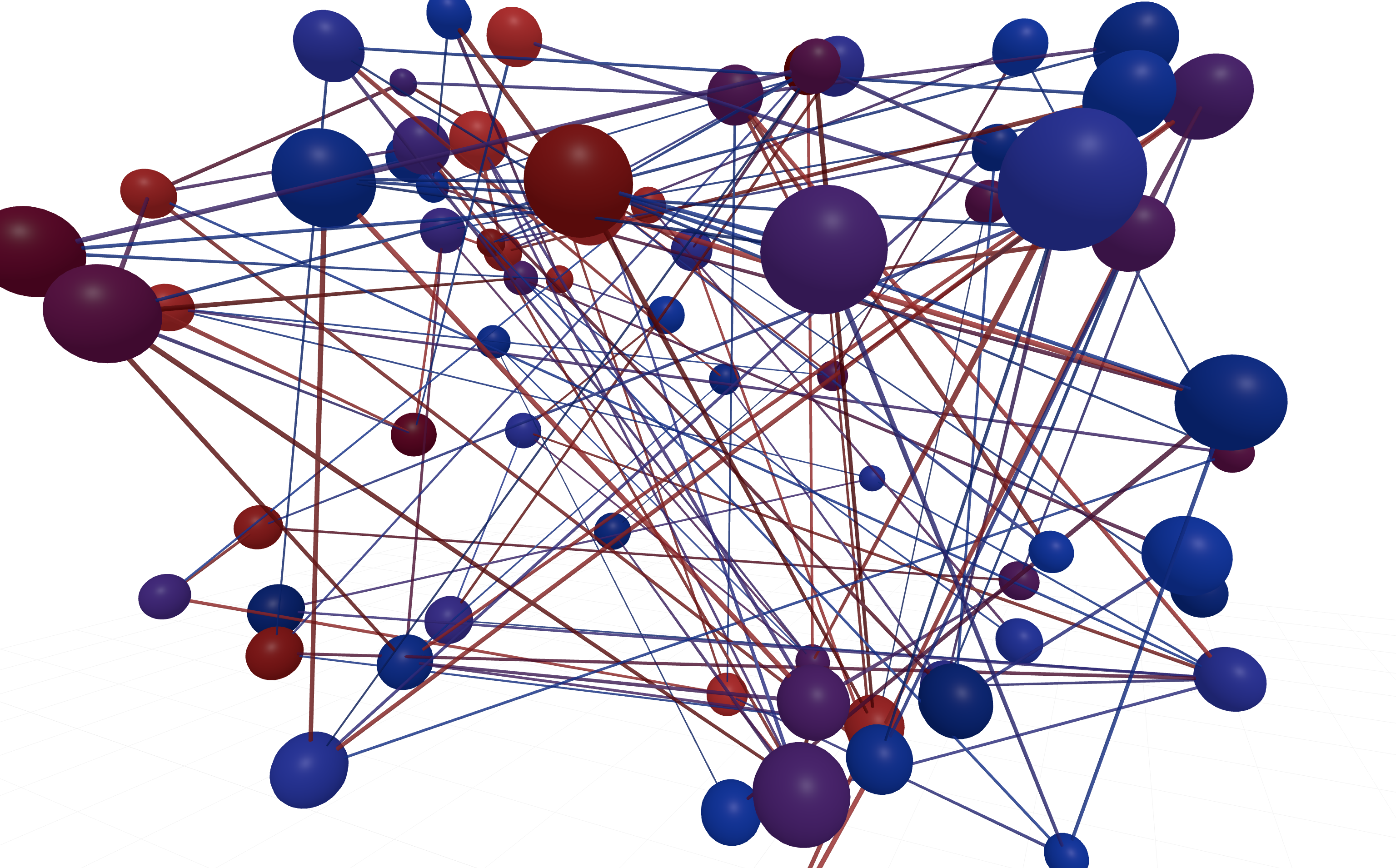}
    \end{minipage}
    
    \vspace{0.5em} 
    
    \begin{minipage}{0.49\textwidth}
        \centering
        \includegraphics[width=\linewidth]{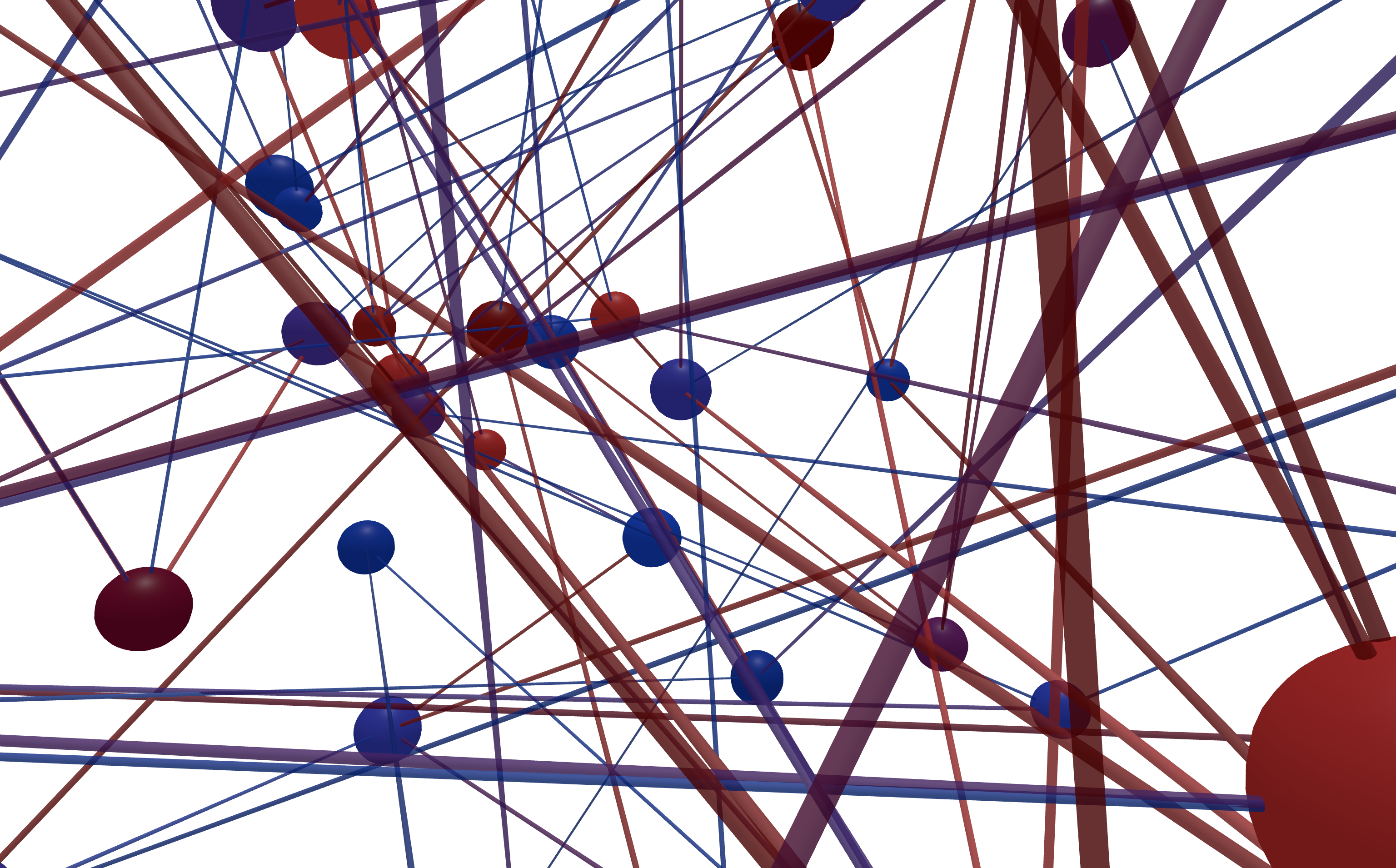}
    \end{minipage}%
    \hfill%
    \begin{minipage}{0.49\textwidth}
        \centering
        \includegraphics[width=\linewidth]{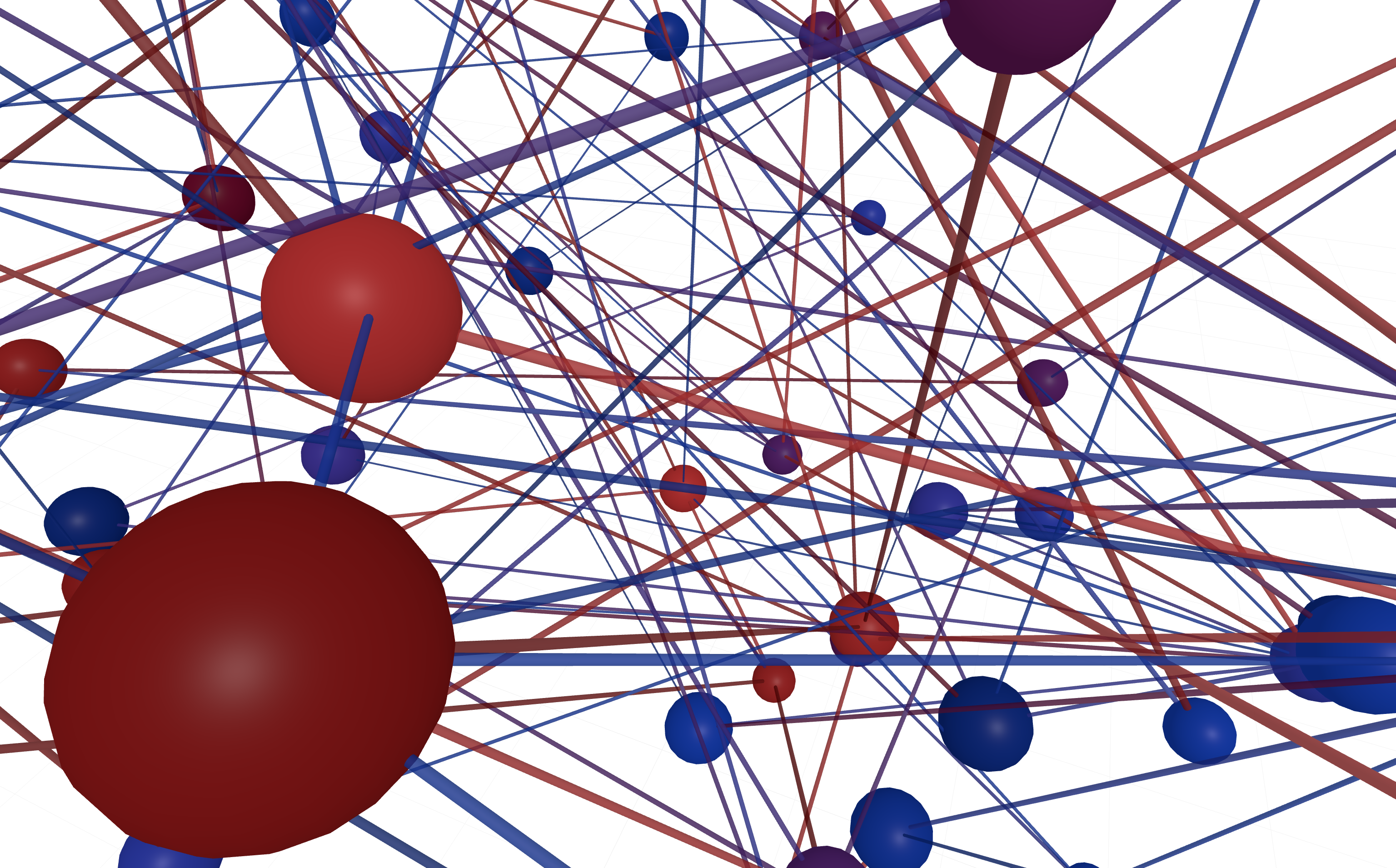}
    \end{minipage}
    
    \caption{Four snapshots of the Mirrored Personal Graph (MPG) visualization exported from the H3LIX/LAIZA Proof-of-Concept interface. Each snapshot corresponds to a system state at time $t$ represented in the quantum-inspired state space. Node size encodes relative activation/weight (salience) of a cognitive/managerial entity in the current state, while node and edge color encode strength on a cold-to-warm scale (dark/cold blue $\rightarrow$ vivid blue $\rightarrow$ dark red $\rightarrow$ bright red), with warmer colors indicating higher activation/stronger relations. The graph provides an interpretable projection of a non-collapsed ambiguity state in which the set of possible scenarios is maintained rather than prematurely reduced.}

\end{figure}

\paragraph{Coherence expectation via Hamiltonian prior dynamics.}
Between observation updates, H3LIX/LAIZA generated a prior expectation of how the state should evolve using a Hermitian operator $\hat{H}_t$ constructed from the MPG structure (\cite{Mathiesen2025b}). In a non-limiting formulation, the Hamiltonian was defined as
\begin{equation*}
\hat{H}_t
=\sum_{v_i,v_j\in V} J_t(v_i,v_j)\,\ket{v_i}\bra{v_j}
+\sum_{v_i\in V} h_t(v_i)\,\ket{v_i}\bra{v_i},
\label{eq:H_case}
\end{equation*}
where the complex couplings $J_t(v_i,v_j)$ were derived from edge-level information (and constrained so that $\hat{H}_t$ is Hermitian), while the local terms $h_t(v_i)\in\mathbb{R}$ were derived from node-level metrics. Informally, $\hat{H}_t$ encoded how activation should propagate across the graph given current relations (edges) and node characteristics. This yielded a predicted prior state via unitary evolution. When new signals arrived, the system reconstructed an updated state from refreshed metrics. The mismatch between the predicted and updated states was treated as a coherence/divergence cue, operationalizing instability in the evolving interpretation.

\begin{figure}[h!]
    \centering
    
    \includegraphics[width=\linewidth]{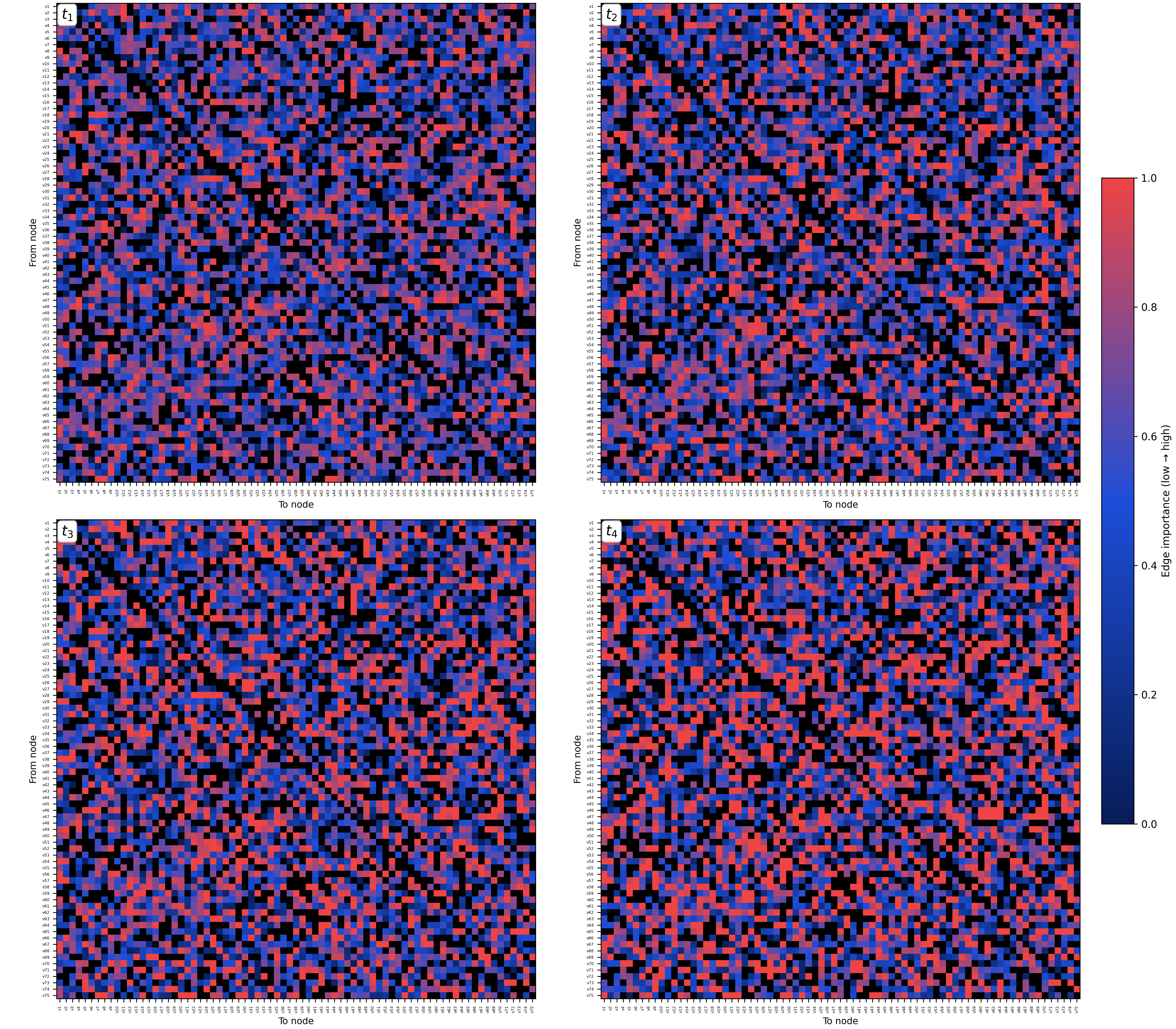}
    
    \caption{Four consecutive snapshots of the H3LIX/LAIZA Mirrored Personal Graph (MPG) edge-importance matrix. Rows and columns correspond to MPG nodes; each colored cell represents a directed edge between two nodes, with color encoding the edge importance/weight. Black cells indicate absent edges. The same set of nodes and candidate edges is retained across panels; the visible color shifts across 
    illustrate the system’s evolving quantum-inspired state representation over time, in which changing relational weights reflect concurrent, non-collapsed interpretive configurations.}

\end{figure}

\paragraph{Divergence, rogue variable detection, and ablation validation.}
H3LIX/LAIZA computed a nonnegative divergence signal $\epsilon_t$ that increased when the predicted evolution and the observation-updated state repeatedly disagreed and/or internal uncertainty indicators remained elevated, capturing situations that remained operationally ``workable'' yet resisted stable interpretation. Over rolling time windows $\mathcal{T}$, high-divergence periods were aggregated into an error-weighted operator
\begin{equation*}
\hat{O}_\epsilon=\frac{1}{Z_\epsilon}\sum_{t\in \mathcal{T}}\epsilon_t\,\ket{\Psi_t}\bra{\Psi_t},
\quad
Z_\epsilon=\sum_{t\in \mathcal{T}}\epsilon_t,
\label{eq:Oeps-case}
\end{equation*}
whose dominant directions highlighted the graph structures most consistently associated with interpretive instability. From this structure, H3LIX/LAIZA identified a candidate ``rogue segment'' $S\subseteq V$ (high-loading nodes) and validated it via ablation logic by testing whether removing the influence of $S$ reduced average divergence over the window. In this case, the dominant segment linked coordination delays, guarded communication about long-term plans, and unresolved IP-boundary issues as a recurring source of mismatch, without implying wrongdoing or classifying intent.

\medskip

Upon detecting persistence, the system surfaced the ambiguity in neutral terms and supported parallel preparedness actions (e.g., legal/patent and exit-readiness) without forcing interpretive closure.

\section{Discussion}

This study was designed not only as a theoretical contribution, but also as an empirical proof of concept for the H3LIX/LAIZA human–AI augmented symbiotic intelligence system and its underlying patented process for Systems and Methods for Quantum-Inspired Rogue Variable Modelling, Human-in-the-Loop Decoherence, and Collective Cognitive Inference in Human–AI Symbiotic Systems (\cite{Mathiesen2025b}). As such, the discussion integrates conceptual implications for management theory with evidence of practical viability, demonstrating how the proposed mechanism operates under real organizational conditions characterized by sustained ambiguity.

\subsection{Ambiguity as a Non-Collapsed State and Managerial Advantage}

A central insight emerging from the case is the strategic value of treating ambiguity as a non-collapsed cognitive state rather than as a failure of decision-making. The H3LIX/LAIZA system operationalized this principle by maintaining multiple plausible interpretations over an extended period, allowing the organization to act without prematurely committing to a single explanatory frame. The case demonstrates that ambiguity, when explicitly represented and managed, can generate temporal and strategic advantage, enabling preparedness without escalation. This finding extends sensemaking theory by showing how ambiguity can be stabilized and engaged productively through human–AI symbiosis rather than resolved through interpretive shortcutting.

\subsection{Weak Signals, Rogue Variables, and Scenario-Based Preparedness}

The proof-of-concept character of the study is particularly visible in the way weak signals were operationalized under conditions of persistent ambiguity. Rather than treating weak signals as isolated anomalies or early indicators that need to be immediately interpreted, the QRVM-based rogue variable detection mechanism identified configurations of signals that reflected sustained interpretive instability within the organization. These rogue variables did not just point to data irregularities; they revealed situations in which existing cognitive, organizational, or managerial models were no longer sufficient to assign stable meaning to incoming signals.

\medskip

By detecting such breakdowns in interpretive coherence, the system shifted the managerial response from predictive certainty to scenario-based preparedness. Instead of converging prematurely on a single forecast or recommendation, decision-makers were supported in maintaining parallel readiness for multiple, mutually exclusive future trajectories. In this way, weak signals were not “resolved” too early but preserved as indicators of emerging change that required ongoing sensemaking.

\medskip

This operationalization demonstrates how weak signal theory can move beyond a primarily conceptual role and be embedded in concrete decision-support mechanisms, while still respecting the inherent ambiguity of VUCA environments. Importantly, the approach avoids premature closure by design, enabling organizations to engage with ambiguity as a manageable cognitive condition rather than a problem to be eliminated through forced interpretation.

\subsection{Human-in-the-Loop Decoherence as Ethical and Governance Control}

The case further validates human-in-the-loop decoherence as a governance mechanism, not a technical afterthought. Once the system detected that internal inference had reached its limits, autonomous interpretation was suspended, and responsibility was explicitly returned to human leadership. This design choice aligns with responsible AI principles and high-reliability organizing, reinforcing ethical restraint under uncertainty. As a proof-of-concept, the study demonstrates that such restraint is not only normatively desirable but operationally effective, preventing both overreaction and inertia during prolonged ambiguity.

\subsection{Organizational Learning Through Ambiguity and Collective Inference}

Finally, the case illustrates how the organizational memory and collective inference components of the patented process function in practice. By capturing and retaining ambiguous episodes as structured learning events, the system enabled early preparation—most notably patent development—without requiring certainty about future outcomes. When ambiguity eventually collapsed, no additional time or reactive effort was needed. This finding supports the argument that learning from ambiguity before resolution constitutes a critical dynamic capability for organizations operating in VUCA environments.

\subsection{Conclusions}

This article advances management science by integrating VUCA theory, weak signal theory, and human–AI symbiosis into a coherent framework for ambiguity management, while simultaneously serving as a proof of concept for the H3LIX/LAIZA system and the patented process for Quantum-Inspired Rogue Variable Modelling, Human-in-the-Loop Decoherence, and Collective Cognitive Inference in Human–AI Symbiotic Systems (\cite{Mathiesen2025b}).

\medskip

Through an in-depth case study, the research demonstrates that ambiguity can be operationalized as a detectable, representable, and governable organizational state. The H3LIX/LAIZA system did not seek to replace managerial judgment or accelerate decision-making; instead, it enhanced timing, preparedness, and ethical control. By preserving ambiguity until it collapsed naturally and supporting parallel scenario preparation—particularly through early patent protection—the system enabled decisive action without delay, disruption, or conflict once clarity emerged.

\medskip

The study makes three principal contributions. First, it reframes ambiguity as a first-class construct in management research, extending beyond decision outcomes to pre-decisional cognitive dynamics. Second, it provides an operational extension of weak signal theory through the concept of rogue variables, demonstrating how early interpretive breakdown can be detected and managed systematically. Third, it contributes to the literature on responsible AI by empirically validating a human–AI symbiotic design in which ethical restraint, human accountability, and organizational resilience are structurally embedded.

\medskip

From a practical perspective, the proof-of-concept shows that organizations operating in VUCA environments benefit not from faster resolution, but from earlier preparedness under uncertainty. Systems that support ambiguity management—rather than eliminate ambiguity—can reduce strategic risk, protect critical assets, and preserve trust even when intentions remain unclear for extended periods.

\medskip

As a proof of concept based on a single organizational case, the study does not claim generalizability across contexts. Future research should examine the application of QRVM-based ambiguity management across multiple organizations, industries, and cultural settings, as well as explore quantitative outcomes related to resilience, decision quality, and governance effectiveness. Longitudinal and comparative studies could further assess how collective cognitive inference scales in larger populations.

\medskip

In an era where uncertainty is structural rather than episodic, the ability to manage ambiguity without collapsing it prematurely may become a defining organizational capability. This study suggests that human–AI symbiotic systems such as H3LIX/LAIZA—designed explicitly for sensemaking rather than certainty—offer a viable and ethically grounded path forward. As both a theoretical contribution and a proof-of-concept, the findings underscore that the future of management lies not in eliminating ambiguity, but in learning how to inhabit it productively.

\printbibliography

\end{document}